\def\year{2020}\relax
\documentclass[letterpaper]{article} 
\usepackage{aaai20}  
\usepackage{times}  
\usepackage{helvet} 
\usepackage{courier}  
\usepackage[hyphens]{url}  
\usepackage{graphicx} 
\usepackage{graphicx}  
\usepackage{amsmath}
\usepackage{amsfonts}
\usepackage{amssymb}
\usepackage{booktabs}
\usepackage{multirow}
\usepackage{subfigure}
\newcommand{\our}{\mbox{PIGAT}}

\title{Pairwise Interactive Graph Attention Network for Context-Aware Recommendation}
\author{Yahui Liu\textsuperscript{\rm 1}, Furao Shen\textsuperscript{\rm 1}, Jian Zhao\textsuperscript{\rm 2} \\ 
\textsuperscript{\rm 1} Department of Computer Science and Technology, Nanjing University, Nanjing, China \\ 
\textsuperscript{\rm 2} School of Electronic Science and Engineering, Nanjing University, Nanjing, China\\ 
liuyahui@smail.nju.edu.cn, \{frshen, jianzhao\}@nju.edu.cn
}
 
\begin{document}

\maketitle

\begin{abstract}
Context-aware recommender systems (CARS), which consider rich side information to improve recommendation performance, have caught more and more attention in both academia and industry. 
How to predict user preferences from diverse contextual features is the core of CARS. 
Several recent models pay attention to user behaviors and use specifically designed structures to extract adaptive user interests from history behaviors. However, few works take item history interactions into consideration, which leads to the insufficiency of item feature representation and item attraction extraction. From these observations, we model the user-item interaction as a dynamic interaction graph (DIG) and proposed a GNN-based model called \underline{P}airwise \underline{I}nteractive \underline{G}raph \underline{At}tention Network ({\our}) to capture dynamic user interests and item attractions simultaneously. {\our} introduces the attention mechanism to consider the importance of each interacted user/item to both the user and the item, which captures user interests, item attractions and their influence on the recommendation context. Moreover, confidence embeddings are applied to interactions to distinguish the confidence of interactions occurring at different times. Then more expressive user/item representations and adaptive interaction features are generated, which benefits the recommendation performance especially when involving long-tail items. We conduct experiments on three real-world datasets to demonstrate the effectiveness of {\our}.
\end{abstract}

\section{Introduction}
Recommender systems (RS), aim to discover the preferred items for potential users, play an increasingly important role in practical applications such as E-commerce and social media. Typically, the core of recommendation systems is to predict user preference precisely, where the preference is usually reflected in rating, clicking, consuming and other user behaviors. When predicting, rich side information such as user profile, item profile and user behaviors is also available beyond the essential user ID and item ID. Context-aware recommendation systems (CARS) are designed to address these highly sparse categorical features to predict user preference more accurately, which has attracted widespread attention in both academia and industry \cite{DBLP:conf/icdm/Rendle10,DBLP:conf/recsys/Cheng0HSCAACCIA16,DBLP:conf/icdm/QuCRZYWW16,DBLP:conf/kdd/LianZZCXS18}.

To obtain the goal of predicting user preference, it is critical to capture user interests and item attractions from numerous features and figure out whether they match each other.
Recently, a series of prediction models pay specific attention to user interest representation by finely dealing with user history behaviors \cite{DBLP:conf/kdd/ZhouZSFZMYJLG18,DBLP:conf/aaai/ZhouBSLZCG18,DBLP:conf/aaai/ZhouMFPBZZG19,DBLP:conf/ijcai/FengLSWSZY19}. However, most CARS models pay less attention to extract item attractions from item historical interaction log. With the historical interaction log, people most likely to be attracted by the item could be captured, which benefits to measure the attraction of the item to a particular user and to enrich item features. These advantages greatly improve the recommendation performance especially for unpopular items since in this way the connections between unpopular items and their interacted users are established, so that more expressive item feature representations could be characterized.
Moreover, both user interests and item attractions are dynamic, interactions occurring at different times have different confidence in representing user interests or item attractions. Thus, it should be taken into account to distinguish the confidence of different interactions.
\begin{figure}[t]
\centering
\includegraphics[scale=0.6]{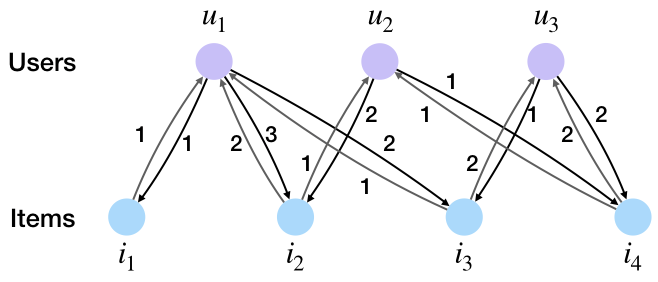}
\caption{An illustration of the dynamic user-item interaction graph. Labels indicate the interaction   order.}
\label{figure-graph}
\end{figure}

Motivated by above observations and inspired by recent developments of graph neural networks (GNN) \cite{DBLP:conf/nips/VaswaniSPUJGKP17,DBLP:conf/iclr/KipfW17,DBLP:conf/iclr/VelickovicCCRLB18} which has the ability to generate hidden representations of graph nodes or the whole graph, we first model the dynamic \mbox{user-item} interactions as dynamic user-item interaction graph (DIG), and then propose \underline{P}airwise \underline{I}nteractive \underline{G}raph \underline{At}tention Network ({\our}) to make full use of dynamic user-item interaction information and improve the prediction performance. As shown in Figure~\ref{figure-graph}, DIG is a directed bipartite graph, each node represents a user or an item, and each directed edge represents an \mbox{user-item} interaction with a label indicates the interaction order in term of the head.

With DIG, our {\our} uses the attention mechanism to capture two types of the importance of each interacted user or item: the importance to the head and the importance to the recommendation context. Considering these importance measurements together, {\our} takes the weighted sum pooling to obtain the interactive head representation as well as adaptive interaction representation. Furthermore, {\our} introduces confidence embeddings to distinguish the confidence of interactions occurring at different times. Hence in our architecture, important interactions have greater impacts on the hidden representations and the importance is measured under multiple criteria, which brings improvement of the representation ability of {\our}.

The main contributions of this work are summarized as follows:
\begin{itemize}
\item We highlight the importance of the \mbox{user-item} interactions and propose a dynamic interaction graph to represent the dynamic interactions between users and items, which can also be taken as a generalization of user historical behaviors.
\item We propose a GNN-based architecture that extracts expressive interactive representations from the interaction graph to improve the performance of content-aware recommendation especially for recommendation involving unpopular items.
\item We employ confidence embedding to distinguish the interactions with different orders and design a novel initialization approach to make the embedding more trainable and effective.
\item Our model achieves \mbox{state-of-the-art} performance in experiments on three \mbox{real-world} datasets and significantly outperforms other comparative models in context-aware recommendation task.
\end{itemize}

\section{Preliminaries}
In this section, we introduce the related concepts and works to our {\our}, which includes \mbox{context-aware} recommendation, \mbox{graph-based} recommendation and \mbox{long-tail} item recommendation.

\subsection{Context-Aware Recommendation}
CARS models consider rich categorical side information besides basic user ID and item ID. Typically, each categorical feature is associated with an embedding to expressively represent the feature, and then complex feature combinations are learned from embeddings. Traditional models use fixed functions to model feature combinations. For example, FM \cite{DBLP:conf/icdm/Rendle10} uses the inner-product to model the combination of each pair of features. Modern models often share the \mbox{Embedding\&MLP} paradigm, in which feature combinations are learned through the MLP network. Such deep models capture complex feature combinations and bring significant improvements in the recommendation performance. \mbox{Deep Crossing} \cite{DBLP:conf/kdd/ShanHJWYM16} concatenates all the embeddings and use residual units to learn feature combinations. \mbox{Wide\&Deep} \cite{DBLP:conf/recsys/Cheng0HSCAACCIA16} combines munually designed features and MLP generated features. DeepFM \cite{DBLP:conf/kdd/LianZZCXS18} combines \mbox{low-order} and \mbox{high-order} feature combinations. 

Recently, attention mechanism \cite{DBLP:journals/corr/BahdanauCB14,DBLP:conf/nips/VaswaniSPUJGKP17} is introduced from Neural Machine Translation (NMT) field to learn the importance of different feature representations. AFM \cite{DBLP:conf/ijcai/XiaoY0ZWC17} follows attention mechanism to distinguish the importance of different feature combinations. More models take effort on the user behavior representation, which improves the simply \mbox{fixed-size} representation used in YoutubeNet \cite{DBLP:conf/recsys/CovingtonAS16}. DIN \cite{DBLP:conf/kdd/ZhouZSFZMYJLG18} adaptively considers the relative importance of each user behavior to the candidate item. ATRank \cite{DBLP:conf/aaai/ZhouBSLZCG18} uses the self-attention mechanism to model heterogeneous user behaviors. 

\subsection{GNN-Based Recommendation}
In recommender systems, \mbox{user-item} interactions are often modeled as an undirected bipartite graph, where users and items are represented by two disjoint parts of the graph and each edge represents the interaction between its endpoints. Inspired by the recent progress in GNN, \mbox{GC-MC} \cite{DBLP:journals/corr/BergKW17} applies the graph convolution network (GCN) \cite{DBLP:conf/iclr/KipfW17} on \mbox{user-item} interaction graph to capture direct \mbox{user-item} relationship, NGCF \cite{DBLP:conf/sigir/Wang0WFC19} builds \mbox{GNN-based} embedding propagation layers to capture collaborative signal through the \mbox{high-order} connectivity. Such models benefit a lot from the strong node representation ability of GNN. However, they mainly focus on the mbox{inner-graph} feature representations and lack the ability to capture the relationship between dynamic interactions and recommendation context.

In this work, we modify the definition of the original interaction graph to satisfy the dynamic setting. As shown in Figure~\ref{figure-graph}, the dynamic \mbox{user-item} interaction graph (DIG) is a directed bipartite graph $G = (V_u, V_i, A)$. User part $V_u$ contains all the user nodes and item part $V_i$ contains all the item nodes. Directed edge set $A$ consists all the directed edges of the form $(h, t, o)$, where $h$ is the head and $t$ is the tail such that $h$ and $t$ are in different parts, label $o$ indicates the interaction order in terms of $h$.
Earliest interaction is labeled $1$ and later is labeled $2, 3, \ldots$. We define the ordered neighbors of node $v$ as the sequence of node reached directly from $v$ sorted by the order of corresponding interactions, that is
\begin{equation}
\label{eq-neighbor}
N_v = (v_1, v_2, \ldots, v_L) \text{ s.t. } \forall 1 \leq l \leq L, (v, v_l, l) \in A,
\end{equation}
where $L$ is the total number of interactions of $v$. Each node in $N_v$ is called a neighbor of $v$. $G$ changes over time by inserting new nodes in $V_u$ or $V_i$ and inserting new interactions into $E$. We denote $G^t$ as the DIG at time $t$. 

\subsection{Long-Tail Item Recommendation}
In \mbox{real-world} recommender systems, only a small number of items have rich interactions whereas the remaining majority have insufficient interactions \cite{anderson2006long,DBLP:journals/pvldb/YinCLYC12}, i.e. items lie in the \mbox{long-tail} distribution. Such majority items are called the \mbox{long-tail} items.
\cite{DBLP:journals/pvldb/YinCLYC12} first proposes the \mbox{long-tail} item recommendation problem, we give the \mbox{DIG-based} \mbox{context-aware} recommendation version of this problem as follows:

Given a DIG $G = (V_u, V_i, A)$ and a query instance $(u, i)$ where $i$ lies in long-tail distribution, predict the probability that user $u$ prefers item $i$.

\section{Model}
\begin{figure*}[ht]
\centering
\includegraphics[scale=0.5]{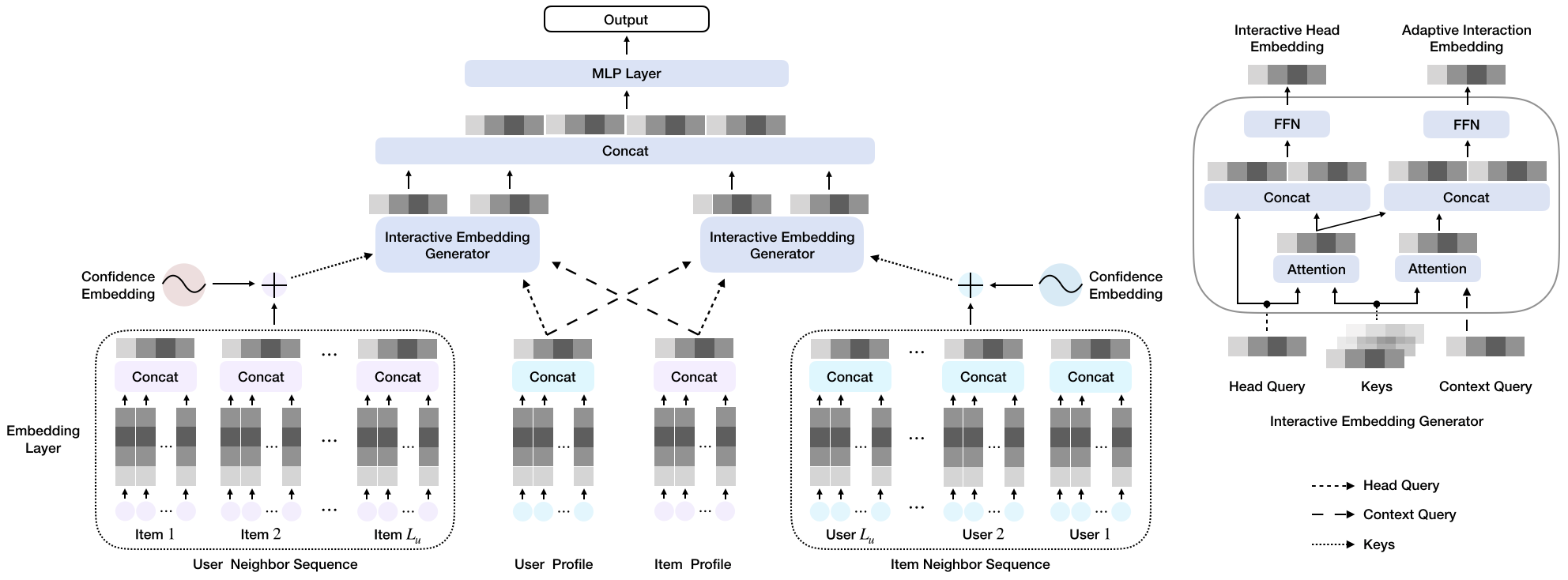}
\caption{Model Architecture.}
\label{figure-model}
\end{figure*}

In this section, we introduce the proposed {\our} in detail, the architecture of which is illustrated in Figure~\ref{figure-model}. {\our} is composed of four parts: (1) an embedding layer that transforms sparse features into dense embedding vectors; (2) a group of confidence embeddings to distinguish the confidence of the interaction neighbors in different positions in the sequence; (3) an interactive embedding generator that generate both the interactive head embeddings and adaptive interaction embeddings; (4) final multilayer perceptron (MLP) layer that predicts the probability that the user prefers the item.  In the rest of this section, we will elaborate these three parts.

\subsection{Feature Representation}
In {\our}, the data we used consist of four groups of categorical features: User Profile, Item Profile, User Neighbor Sequence, and Item Neighbor Sequence. User Neighbor Sequence contains the sequence of item profiles corresponding to the ordered neighbors of the user and Item Neighbor Sequence contains the sequence of user IDs corresponding to the ordered neighbors of the item, where ordered neighbors are the nodes directly reached from the user/item in DIG in the order of corresponding interaction's order as defined in Equation~\ref{eq-neighbor}. Each group of categorical features is represented by a \mbox{high-dimensional} sparse binary feature via one-hot embedding or \mbox{multi-hot} embedding. An example is illustrated as follows:
\[
\underbrace{\left[ 1, 0, \ldots, 0 \right]}_{\mbox{user\_id=0}} \quad \underbrace{\left[ 0, 1, 0, \ldots, 0 \right]}_{\mbox{item\_id=1}} \quad \underbrace{\left[ 0, 0, 1, 0, \ldots, 0 \right]}_{\mbox{item\_cate\_id=Comedy}} 
\]
\[
\underbrace{\left[ 1, 0, 1, 0, \ldots, 0 \right]}_{\mbox{user\_neighbors=\{0, 2\}}} \quad \underbrace{\left[ 0, 1, 0, 1, 0, \ldots, 0 \right]}_{\mbox{item\_neighbors=\{1, 3\}}}
\]

\subsection{Embedding Layer}
In embedding layer, \mbox{high-dimensional} sparse features are transformed into \mbox{low-dimensional} dense vectors by looking up embedding tables. Specifically, we build user embedding table $\mathbf{E}_u = [\mathbf{e}_{u_1}, \mathbf{e}_{u_2}, \ldots, \mathbf{e}_{u_{N_u}}]\in \mathbb{R}^{H_u \times N_u}$
to represent User Profile, where $H_u$ is the embedding size of User Profile, $N_u$ is the total number of categorical features in User Profile and $\mathbf{e}_{u_k} \in \mathbb{R}^{H_u}$ is an embedding vector with dimensional of $H_u$. Item embedding table $\mathbf{E}_i \in \mathbb{R}^{H_i \times N_i}$ is built in a similar way.

Given an input vector $\mathbf{x} \in \{0, 1\}^{N}$ and embedding table $\mathbf{E} = [\mathbf{e}_1, \mathbf{e}_2, \ldots, \mathbf{e}_N] \in \mathbb{R}^{H \times N}$,  the embedding of $\mathbf{x}$ is $\mathbf{e} = [\mathbf{e}_{k_1}, \mathbf{e}_{k_2}, \ldots, \mathbf{e}_{k_n}]$, where $j \in \{k_1, k_2, \ldots, k_n\}$ if and only if $x_j = 1$. In this way, User Profile, Item Profile, User Neighbor Sequence, and Item Neighbor Sequence are represented by the embedding vectors $\mathbf{e}_{u} \in \mathbb{R}^{n_u H_u}$, $\mathbf{e}_{i} \in \mathbb{R}^{n_i H_i}$, $\mathbf{e}_{uns} \in \mathbb{R}^{L_u \times n_i H_i}$ and $\mathbf{e}_{ins} \in \mathbb{R}^{L_i \times H_u}$, respectively, where $n_u$ ($n_i$) is the number of categorical features in User Profile (Item Profile) and $L_u$ ($L_i$) is the count of neighbors of the user (item). Note that User Neighbor Sequence shares the same embedding table with Item Profile and Item Neighbor Sequence shares the same embedding table with User Profile, which enables {\our} to integrate the graph information with contextual features.

\subsection{Confidence Embedding}
In recommender systems, recent interactions are usually more reflective of user preference than previous interactions, which should play a more credible role in the recommendation process. To distinguish the confidence of interactions occurring at different time, we introduce the confidence embedding into {\our}. In this work, we initialize the confidence embedding with the following equation:
\begin{equation}
\label{eq-ce}
CE_{(l, i)} = \exp(l - L - 1) \cos((i - 1) \pi / H),
\end{equation}
where $l$ indicates the order of the interaction, $i$ indicates the index of the unit in the embedding, $L$ is the total number of interactions and $H$ is the dimension of the embedding. Note that for a particular interaction, the confidence embedding of it forms a cosine curve with length $\pi$; for a particular embedding index, the embedding value is exponential decay according to the time-reverse order of interactions. Since interactions are represented by the ordered neighbors in our model, then given the input interaction neighbor embedding sequence, the confidence embedding with the same embedding dimension is added to it before computing the attention coefficients.

Our confidence embeddings share the similar idea as positional encodings in NMT task \cite{DBLP:conf/icml/GehringAGYD17,DBLP:conf/nips/VaswaniSPUJGKP17}, which is to make use of the order of sequence. However, we use exponential scalar to model the decay of the interaction confidence instead of just distinguish the different position in the sequence.

\subsection{Interactive Embedding Generator}
{\our} aims to capture dynamic and adaptive user interests and the item attractions from the DIG. Thus in contrast with traditional models like \cite{DBLP:conf/kdd/ShanHJWYM16,DBLP:conf/recsys/Cheng0HSCAACCIA16} in which embeddings are directly fed into the MLP network, {\our} uses the interactive embedding generator to generate more effective embeddings before MLP layer.

As shown in Figure~\ref{figure-model}, interactive embedding generator consists of two components: (1) a pairwise attention layer to capture the interactive relationship between ordered neighbors and the head, and the adaptive relationship between user interests (item attractions) and recommendation context; (2) an integrate layer to generate the final representation of head node features and adaptive interaction embeddings. 

\subsubsection{Pairwise Attention Layer}
In recommender systems, \mbox{user-item} interactions reflect user interests and item attractions, which can be used to enhance the primary user/item representation and estimate the relevance to the recommendation context. Based on this fact, we introduce the pairwise attention layer to measure the importance of each interacted neighbor to both the head and the context, and then perform weighted sum pooling to generate interactive embeddings. Mathematically, after the embedding layer as described above, we obtain $\mathbf{e}_{u}$, $\mathbf{e}_{i}$, $\mathbf{e}_{uns}$ and $\mathbf{e}_{ins}$ to represent User Profile, Item Profile, User Neighbor Sequence and Item Neighbor Sequence, respectively. We then take these embeddings as the input embeddings of the pairwise attention layer, and compute the attention coefficients
\begin{equation}
\begin{split}
a_{ui}^{(l)} & = \operatorname{softmax}(\operatorname{FFN}_{ui}(\mathbf{e}_{u}, \mathbf{e}_{uns}^{(l)})), \\
a_{ua}^{(l)} & = \operatorname{softmax}(\operatorname{FFN}_{ua}(\mathbf{e}_{u}, \mathbf{e}_{uns}^{(l)})), \\
a_{ii}^{(l)} & = \operatorname{softmax}(\operatorname{FFN}_{ii}(\mathbf{e}_{u}, \mathbf{e}_{ins}^{(l)})), \\
a_{ia}^{(l)} & = \operatorname{softmax}(\operatorname{FFN}_{ia}(\mathbf{e}_{u}, \mathbf{e}_{ins}^{(l)})),
\end{split}
\end{equation}
where $\cdot^{(l)}$ denotes the embedding of the $l$-th neighbor, $\operatorname{FFN}_{ui}(\cdot)$, $\operatorname{FFN}_{ua}(\cdot)$, $\operatorname{FFN}_{ii}(\cdot)$, and  $\operatorname{FFN}_{ua}(\cdot)$ are four independent feed-forward neural networks (FFNs) with the same structure but different weights and the output of each FFN is normalized through the softmax function:
\begin{equation}
\operatorname{softmax} \left(x_{i}\right)=\frac{\exp \left(x_{i}\right)}{\sum_{j} \exp \left( x_j \right)}.
\end{equation}
$a_{ui}$, $a_{ua}$, $a_{ii}$, $a_{ia}$ are called user interactive weight, user adaptive weight, item interactive weight, item adaptive weight, respectively. Note that for user (item) neighbor sequence, interactive weights characterize the importance of each neighbor to the head, and adaptive weights characterize the importance of each neighbor to the context.
Finally, the obtained attention coefficients are used to calculate the interactive embeddings and adaptive embeddings as follows:
\begin{equation}
\begin{split}
\mathbf{h}_{ui} = \sum_{l = 1}^{L_u} a_{ui} \mathbf{e}_{uns}^{(l)}, & \quad \mathbf{h}_{ua} = \sum_{l = 1}^{L_u} a_{ua} \mathbf{e}_{uns}^{(l)}, \\
\mathbf{h}_{ii} = \sum_{l = 1}^{L_i} a_{ii} \mathbf{e}_{ins}^{(l)}, & \quad \mathbf{h}_{ia} = \sum_{l = 1}^{L_i} a_{ia} \mathbf{e}_{ins}^{(l)},
\end{split}
\end{equation}
where $L_{u}$, $L_{i}$ are the neighbor number as described above.

\subsubsection{Integrate Layer}
After the pairwise attention layer, the outputted interactive embeddings and adaptive embeddings are integrated with original profile embeddings in the integrate layer. Specifically,
the interactive embedding and original profile embedding are concatenated, then single-layer FFNs are used to generate the final interactive embeddings, which can be formulated as follows:
\begin{equation}
\begin{split}
\mathbf{h}_{u} & = \operatorname{LeakyReLU} \left( \mathbf{W}_u [\mathbf{e}_{u} \| \mathbf{h}_{ui}] + \mathbf{b}_u \right), \\
\mathbf{h}_{i} & = \operatorname{LeakyReLU} \left( \mathbf{W}_i [\mathbf{e}_{i} \| \mathbf{h}_{ii}] + \mathbf{b}_i \right),
\end{split}
\end{equation}
where $\mathbf{W}_u \in \mathbb{R}^{n_i H_i \times n_i H_i}$, $\mathbf{W}_i \in  \mathbb{R}^{n_u H_u \times H_u}$ are the weights of FFNs to generate user interactive embedding and item interactive embedding respectively, $\mathbf{b}_u \in \mathbb{R}^{n_i H_i}$, $\mathbf{b}_i \in \mathbb{R}^{n_u H_u}$ are the bias of FFNs, and $\|$ is the concatenation operation. In addition, \mbox{LeakyReLU} activation is applied to the fully-connected layer. Analogously, the outputted interactive embedding and adaptive embedding are concatenated then fed into the FFNs to generate adaptive interaction embeddings $\mathbf{h}'_{u}$ and $\mathbf{h}'_{i}$.

\subsection{MLP Layer}
In MLP layer, the interactive embeddings ($\mathbf{h}_{u}$, $\mathbf{h}_{i}$) and adaptive interaction embeddings ($\mathbf{h}'_{u}$, $\mathbf{h}'_{i}$) generated by the interactive embedding generator are concatenated, and then fed into the final MLP to predict the probability that user prefers the item.

\subsection{Loss Function}
Given an training instance $(\mathbf{x}, y)$, our target is to maximize the predicted probability $\hat{y}=f(\mathbf{x})$ if $y = 1$, otherwise ($y = 0$) is to minimize $\hat{y}$. Since then, we take the negative log-likelihood function as our loss function, which is defined as follows:
\begin{equation}
L = -\frac{1}{n} \sum_{(\mathbf{x}, y) \in D}(y \log f(\mathbf{x})+(1-y) \log (1-f(\mathbf{x}))),
\end{equation}
where $D$ is the training set.

\section{Experiments}
In this section, we perform experiments on three real-word datasets to evaluate the performance of our proposed {\our}. We start by introducing the detailed experimental setup, and then presents the experiment results and analysis. Experiments shows that {\our} outperforms state-of-the-art methods on user preference prediction task.

\subsection{Experimental Setup}
\subsubsection{Datasets}
We conduct experiments on both benchmark datasets and tens-of-millions sized grand challenge datasets to evaluation the effectiveness of our proposed approach. 
\begin{itemize}
\item {\textbf{Amazon}\footnote{http://jmcauley.ucsd.edu/data/amazon/}.}
Amazon dataset is a widely used benchmark dataset \cite{DBLP:conf/www/HeM16,DBLP:conf/sigir/McAuleyTSH15,DBLP:conf/kdd/ZhouZSFZMYJLG18}, which contains product reviews and metadata from Amazon. All users and items in the dataste have at least 5 reviews. Our experiments are conducted on two subsets of Amazon Dataset: Books and Electronics. We select the ID of the reviewer as the User Profile, select the ID and categories of the product as the Item Profile, take the reviewed product as User Neighbors and take the reviewing user as Item Neighbors. Furthermore, we label the instances with overall rating above 3 as positive and the rest as negative.
\item {\textbf{Byte-Recommend}\footnote{https://biendata.com/competition/icmechallenge2019/data/}.}
Byte-Recommend dataset is a large public grand challenge dataset, which contains of tens of thousands of different users and millions of different videos.  We use user\_id and device\_id as User Profile, use item\_id and author\_id as Item Profile, take the watched videos as User Neighbors and take the watching user as Item Neighbors. We directly use the finish indicator (indicating whether the user finishes watching the video) in the dataset as the label.
\end{itemize}
The statistics of all above datasets are shown in Table~\ref{table-datasets}. Note that in Byte-Recommend dataset, item density is much lower than other datasets, which leads to more serious long-tail problem.
\begin{table*}[ht]
\centering
\caption{Statistics of datasets.}
\label{table-datasets}
\begin{tabular}{cccccc}
\toprule
Dataset & Instances & Users & Items & avg. \# of Users & avg. \# of Items \\
\midrule
Electronics & 1,689,188 & 192,403 & 63,001 & 8.8 & 26.8 \\
Books & 8,898,041 & 603,668 & 367,982 & 14.7 & 24.2 \\
Byte-Recommend & 19,622,340 & 73,974 & 4,122,689 & 277.5 & 5.3 \\
\bottomrule
\end{tabular}
\end{table*}

\subsubsection{Competitors}
We compare our model with the following models to evaluate the performance:
\begin{itemize}
\item \textbf{FM} \cite{DBLP:conf/icdm/Rendle10}
Factorization machine (FM) is a classical context-aware recommendation model, which captures the feature interaction through the \mbox{inner-product}.
\item \textbf{YoutubeNet} \cite{DBLP:conf/recsys/CovingtonAS16}
YoutubeNet is a classical model following the \mbox{Embedding\&MLP} paradigm. 
\item \textbf{DeepFM} \cite{DBLP:conf/ijcai/GuoTYLH17}
DeepFM models \mbox{low-order} relationship and \mbox{high-order} relationship between features simultaneously and shares embeddings between two components.
\item \textbf{DIN} \cite{DBLP:conf/kdd/ZhouZSFZMYJLG18}
DIN is a \mbox{state-of-art} model for context-aware recommendation, which use attention mechanism to learn the relationship between user behaviors and the candidate item. We take the user interactive neighbors as user behaviors and take the item interactive neighbors as common context features.
\item \textbf{GC-MC} \cite{DBLP:journals/corr/BergKW17}
\mbox{GC-MC} is a \mbox{state-of-art} \mbox{GCN-based} model. We take the user profile and the item profile as side information.
\end{itemize}
For FM, we conduct experiments with and without interaction neighbor sequence (refer as FM$-$) and for YoutubeNet, we conduct experiments with and without item interaction neighbor (refer as YoutubeNet$-$) to verify the effectiveness of the interaction neighbor sequence. In FM and DeepFM, the interaction neighbor sequence are treated as undifferentiated sparse features. In YoutubeNet, the interaction neighbor sequence are tuned into fixed-length embedding through average pooling operation.

\subsubsection{Evaluation Protocols}
We split each dataset into training ($80\%$), validation ($10\%$), and test ($10\%$) sets according to the timeline, where the validation set is to tune hyper-parameters and performance comparisons are taken on the test set. The task is to predict the label in each dataset. We only use last $10$ interactions (with the largest $10$ labels) of users or items with no more than $10$ interactions for all models and all datasets. To be fair, we implement all models in Pytorch and use Adam optimizer \cite{DBLP:journals/corr/KingmaB14} to optimize all models. The embedding size is fixed to $64$ on \mbox{Byte-Recommend} dataset and $128$ on all other datasets for all models. For YoutubeNet, DeepFM, DIN and {\our}, the MLP layer is set to contain three layers with hidden size 80, 40, 1, respectively. The batch size is fixed to $4096$. We set the learning rate decayed by constant rate every 1 or 2 epoch(s), the learning rate is selected in $\{10^{-5}, 5 \times 10^{-5} , 10^{-4}, 5 \times 10^{-4}, 10^{-3}\}$, and the decay rate is selected in $\{0.1, 0.2, 0.5, 1.0\}$. To overcome the overfitting problem, we apply $L_2$ regularization with the coefficient in $\{10^{-2}, 10^{-3}, 10^{-4}, 10^{-5}\}$ and dropout \cite{DBLP:journals/jmlr/SrivastavaHKSS14} with the ratio in $\{0.0, 0.1, 0.2, 0.5\}$ is applied to the input of the MLP layer.

\subsubsection{Metrics}
Area Under ROC Curve (AUC) measures the ranking ability of the model \cite{DBLP:journals/prl/Fawcett06}, which is a widely used metric in \mbox{context-aware} recommendation. It is defines as follows:
\begin{equation}
\operatorname{AUC} = \frac{1}{m^{+} m^{-}} \sum_{\mathbf{x}^{+} \in D^{+}} \sum_{\mathbf{x}^{-} \in D^{-}} R\left(f(\mathbf{x}^{+}), f(\mathbf{x}^{-}) \right),
\end{equation}
where $D^{+}$ is the set of all positive instances with size $m^+$, $D^{-}$ is the set of all negative instances with size $m^-$, $f(\cdot)$ is the prediction model and $R: \mathbb{R} \times \mathbb{R} \rightarrow \mathbb{R}$ is a function to compute the ranking score of the input pair, which is defined as follows:
\begin{equation}
R\left(x_1, x_2 \right) = \mathbb{I}\left(x_1 > x_2 \right) + \frac{1}{2} \mathbb{I}\left(x_1 = x_2 \right),
\end{equation}
where $\mathbb{I}(\cdot)$ is the indicator function.

\subsection{Performance Comparison Results}
\subsubsection{Overall Comparison}
Table~\ref{table-comparison} shows the AUC score on the test set, from which we have the following observations:
\begin{itemize}
\item YoutubeNet performs better when using item neighbor sequence on all three datasets, FM performs better when using interaction neighbor sequences on two larger datasets, which verifies the importance of introducing \mbox{user-item} interactions into recommendation models.
\item All the deep models achieve better performance than FM, which indicates that using \mbox{inner-product} to capture only the \mbox{low-order} feature interactions is insufficient. 
DeepFM outperforms YoutubeNet on Amazon Electronics and \mbox{Byte-Recommend} but underperforms on Amazon Books, which implies that it's insufficient to regard the interaction neighbor sequence same as other features, although complex feature relationship is taken into consideration. DIN improves the performance significantly owe to its specially designed structure to extract user interests. \mbox{GC-MC} captures the relationship between user/item and neighbors in interaction graph, which demonstrates the final representation of user/item features and might limit the model performance for inadequate expressive of User Profile and Item Profile.
\item Our model achieves best performance on all datasets in terms of AUC score, especially on \mbox{Byte-Recommend} Dataset with a large volume of long-tail items. It learns both the importance to the head and the adaptive relationship between interactive neighbors and recommendation context, thereby obtaining more effective interactive embeddings to represent user preference and item attractions.
\end{itemize}
\begin{table}[ht]
\centering
\caption{Performance Coparison on all Datasets.}
\label{table-comparison}
\begin{tabular}{cccc}
\toprule
Model & Books & Electronics & Byte-Rec \\
\midrule
FM$-$ & 0.6596 & 0.6279 & 0.6822 \\
FM & 0.6763 & 0.6183 & 0.6979 \\
YoutubeNet$-$ & 0.7643 & 0.7004 & 0.7310 \\
YoutubeNet & 0.7677 & 0.7014 & 0.7385 \\
DeepFM & 0.7666 & 0.7016 & 0.7391 \\
DIN & \underline{0.7684} & \underline{0.7027} & \underline{0.7392} \\
GC-MC & 0.7668 & 0.7025 & 0.7387 \\
\textbf{\our} & \textbf{0.7694} & \textbf{0.7033} & \textbf{0.7422} \\
\bottomrule
\end{tabular}
\end{table}

\subsubsection{Long-Tail Recommendation Comparison}
\begin{figure*}[ht]
\centering
\subfigure[AUC on Books]{ 
    \label{figure-longtail-books} 
    \includegraphics[scale=0.35]{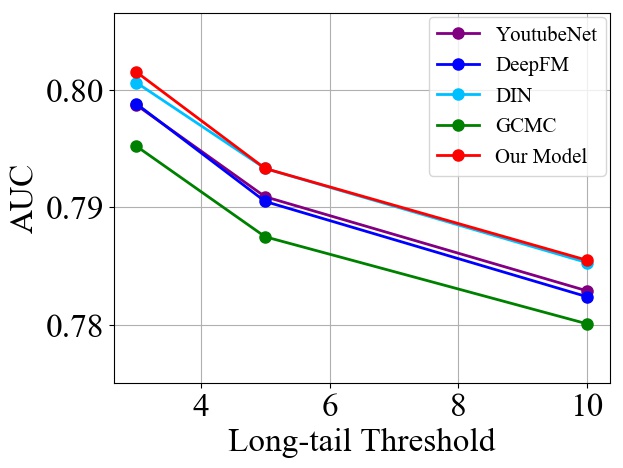}}
\subfigure[AUC on Electronics]{ 
    \label{figure-longtail-electronics}
    \includegraphics[scale=0.35]{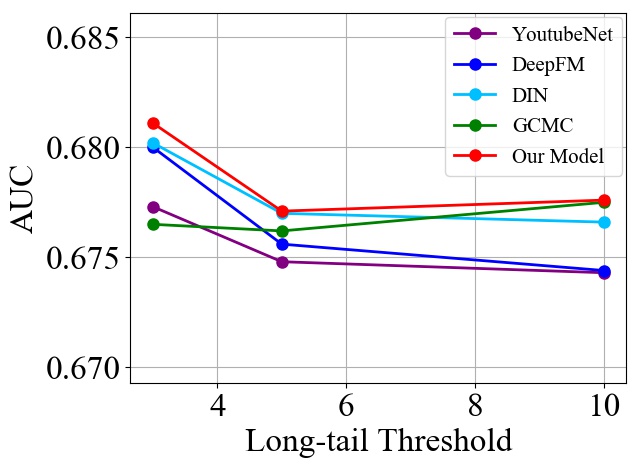}}
\subfigure[AUC on Byte-Recommend]{ 
    \label{figure-longtail-byte}
    \includegraphics[scale=0.35]{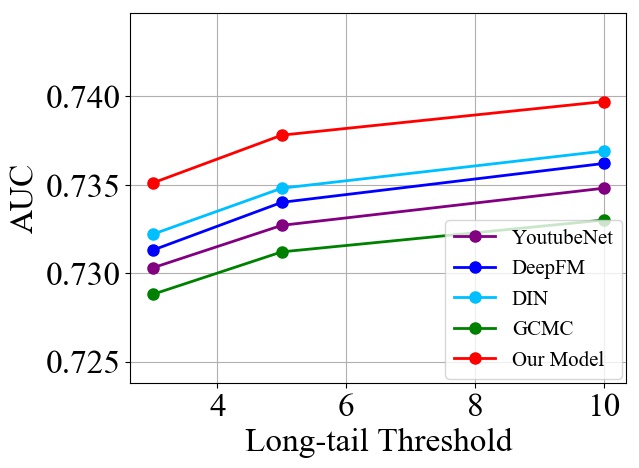}}
\caption{Performance comparison under different long-tail threshold.}
\label{figure-longtail}
\end{figure*}

To verify the effectiveness of {\our} to do \mbox{long-tail} recommendation, we calculate the AUC score on the \mbox{long-tail} subsets which contain items with no more than $k$ neighbors in the training set. The $k$ is called the \mbox{long-tail} threshold and is chosen from $\{3, 5, 10\}$. Results are shown in Figure~\ref{figure-longtail}, we omit the result for FM since other models beat it significantly.

It can be observed that the performance of different models shows a similar trend on each dataset. {\our} achieves the best performance on all three datasets, especially on \mbox{Byte-Recommend} dataset which contains numerous \mbox{long-tail} items. This verifies that the embedding representations generated by our interaction embedding generator improve the \mbox{long-tail} item recommendation indeed. What's more, {\our} improves the performance more significantly when long-tail threshold $k=3$ than larger thresholds, which further verified the ability of {\our} to generate expressive enough representations for items with extremely rare interactions.

\subsection{Study of our Model}

\subsubsection{Effect of Dynamic Interaction Graph Features}
To verify the effectiveness of dynamic interaction graph features, we replace the dynamic edges by static edges representing the latest $10$ interactions in the training set. Table~\ref{table-dynamic} summarizes the results. 
It shows that recommendation performance decreases a lot by using the static interaction graph. This might be caused by the mismatch between the static interactions and dynamic user interests. This verifies the necessity to introduce dynamic interactions into content-aware recommendation systems.
\begin{table}[ht]
\centering
\caption{Effect of dynamic interaction graph features. }
\label{table-dynamic}
\begin{tabular}{cccc}
\toprule
Graph Type & Books & Electronics & Byte-Rec \\
\midrule
Static & 0.7684 & 0.6960 & 0.7234 \\
\textbf{Dynamic} & \textbf{0.7694} & \textbf{0.7033} & \textbf{0.7422} \\
\bottomrule
\end{tabular}
\end{table}

\subsubsection{Effect of Confidence Embedding}
To study the influence of confidence embedding to recommendation performance, we conduct experiments on following variants by replacing our confidence embedding to other structures: removing confidence embedding which is donated as None, using positional embedding as \cite{DBLP:conf/nips/VaswaniSPUJGKP17} which is donated as PE, using fixed confidence embedding as Equation~\eqref{eq-ce} which is donated as FCE, using random initialed learned embedding which is denoted as RCE. Table~\ref{table-CE} summaries the results, from which we observed that 
\begin{itemize}
\item CE achieves best performance, which verifies the effectiveness of our confidence embedding.
\item Using fixed confidence embedding decreases the performance obviously, which implies the variety of interaction confidence might not be able to be generalized by the same rule.
\item FCE outperforms PE in most cases, which indicates our confidence embedding is more suitable to model interaction confidence than positional embedding.
\item RCE sometimes outperforms model with out confidence embedding, but generally underperforms CE, which further verifies the necessary to introduce special designed confidence embedding.
\end{itemize}
\begin{table}[ht]
\centering
\caption{Effect of confidence embedding. }
\label{table-CE}
\begin{tabular}{cccc}
\toprule
Model & Books & Electronics & Byte-Rec \\
\midrule
None & \underline{0.7685} & 0.7009 & 0.7416 \\
PE & 0.7623 & 0.6960 & 0.7407 \\
FCE & 0.7671 & 0.6991 & 0.7414 \\
RCE & 0.7683 & \underline{0.7010} & \underline{0.7419} \\
\textbf{CE} & \textbf{0.7694} & \textbf{0.7033} & \textbf{0.7422} \\
\bottomrule
\end{tabular}
\end{table}

\subsubsection{Effect of Attention Function}
To study how attention functions influence our architecture, we conduct experiments on variants using dot-product, scaled dot-product \cite{DBLP:conf/nips/VaswaniSPUJGKP17}, and FFN as attention functions. As shown in Table~\ref{table-attention}, using FFN generally outperforms using (scaled) dot-product attention and using FFN-3 achieves best performance, which implies the existence of higher-order relationship between graph node representations and context features. 
\begin{table}[ht]
\centering
\caption{Effect of attention function. Dot-product denotes model using dot-product attention, Dot-product-S denotes model using scaled dot-product attention, FFN-$n$ denotes model using $n$ layer(s) FFN as attention function.}
\label{table-attention}
\begin{tabular}{cccc}
\toprule
Model & Books & Electronics & Byte-Rec \\
\midrule
Dot-product & \underline{0.7683} & 0.7007 & 0.7413 \\
Dot-product-S & 0.7682 & 0.7010 & 0.7417 \\
FFN-1 & \underline{0.7683} & \underline{0.7020} & 0.7417 \\
FFN-2 & 0.7682 & 0.7019 & \underline{0.7419} \\
\textbf{FFN-3} & \textbf{0.7694} & \textbf{0.7033} & \textbf{0.7422} \\
\bottomrule
\end{tabular}
\end{table}

\subsubsection{Further Discussion}
Jointly analyze Table~\ref{table-comparison}, Table~\ref{table-CE}, Table~\ref{table-attention} and Figure~\ref{figure-longtail} we have observed that without confidence embedding, our model already outperforms other model in Books and Byte-Recommend, but confidence embedding leads to further improvements of the performance. For Byte-Recommend, replacing either the confidence embedding or the attention function does not affect the model to achieve best performance, which indicates the superiority of the interactive embedding generator framework for \mbox{long-tail} item recommendation.

\section{Conclusion}
In this work, we propose a \mbox{GNN-based} \mbox{context-aware} recommendation model {\our}, which follows attention mechanism to generate expressive representation of \mbox{user-item} interactions and interactive user/item representations. To capture dynamic user interests, we use the dynamic \mbox{user-item} interaction graph rather than a static graph. The key of {\our} is to consider two types of the importance of each interaction neighbor: the importance to the head and the importance to the candidate. With the previous one, a more expressive head node representation can be generated. With both of them, the relationship between user interests, item attractions and recommendation context can be captured. We further apply the confidence embeddings to model the variety of interaction confidence. Experiments on three datasets show that the above considerations improve the model performance significantly especially for long-tail item recommendation. 

\bibliography{ref}
\bibliographystyle{aaai}
\end{document}